\documentclass[%
 reprint,
 superscriptaddress,
 amsmath,amssymb,
 aps,
]{revtex4-1}

\usepackage{graphicx,epsf}
\graphicspath{{Fig/}}
\usepackage{bm}
\usepackage{float}

\usepackage{color}
\newcommand{\blue}{}

\begin{document}

\title[]{Thermoelectric properties of tubular nanowires in the presence of a transverse magnetic field}


\author{Hadi Rezaie Heris}
\affiliation{School of Science and Engineering, Reykjavik University, Menntavegur 1, IS-102 Reykjavik, Iceland}

\author{Movaffaq Kateb}
\affiliation{School of Science and Engineering, Reykjavik University, Menntavegur 1, IS-102 Reykjavik, Iceland}

\author{Sigurdur I. Erlingsson}
\affiliation{School of Science and Engineering, Reykjavik University, Menntavegur 1, IS-102 Reykjavik, Iceland}

\author{Andrei Manolescu}
\affiliation{School of Science and Engineering, Reykjavik University, Menntavegur 1, IS-102 Reykjavik, Iceland}

\begin{abstract}
We calculate the charge and heat current associate with electrons, generated by a temperature gradient and chemical potential difference between two ends of a tubular nanowire of 30\,nm radius in the presence of an external magnetic field perpendicular to its axis. 
We consider a nanowire based on a semiconductor material, and use the Landauer-B\"{u}ttiker approach to calculate the transport quantities. 
We obtain the variation of the Seebeck coefficient ($S$), thermal conductivity ($\kappa$), and the figure of merit ($ZT$), with respect to the temperature up to 20\,K, and with the magnetic field up to 3\,T. In particular we show that the Seebeck coefficient can change sign in this domain of parameters.  In addition $\kappa$ and $ZT$ have oscillations when the magnetic field increases.  These oscillations are determined by the energy spectrum of the electrons. 
\end{abstract}

\maketitle

\section{Introduction}

Thermoelectric materials have attracted considerable attention due to their 
potential applications in electronics
\citep{paulsson2003thermoelectric,weber2006coin,suarez2016designing,suarez2017flexible},
as energy conversion devices
\citep{snyder2011,wood1988materials,fergus2012oxide,nolas1999skutterudites,prete2019thermoelectric},
or 
as components of complex instruments used in medical
science \citep{cosman1990thermometric}. Thermoelectric devices
display interesting properties such as being very reliable because
they do not contain any moving part, being of very small size, and
most importantly, capable of energy harvesting from waste
heat of environment, that makes them very attractive for industry
\citep{dudzinski2014nasa,vining2009inconvenient,snyder2011}.
Semiconductor nanowires are promising candidates for
thermoelectric applications, along with their rich and complex
electrical, optical, and photovoltaics properties
\citep{zhao2004quantum,kateb2013fast,kateb2016,tian2009single}. Nanowires
have played an important role in this research direction due
to their ability to provide efficient thermoelectric elements
with low thermal conductivity and high figure of merit ($ZT$)
\citep{rossella2018measurement,hong2020establishing,diez2020enhanced}. 

{\blue In particular, core-shell nanowires based on III-V semiconductors enable the control of charge, and possibly heat transfer through the specific geometry and shell thickness. With a doped shell and undoped core one can obtain a tubular conductor \cite{Gul2014} with conduction electrons captured inside the shell.  Most often such nanowires have a hexagonal cross section and the charge density peaks at the shell corners \citep{Ferrari09,sitek2016multi,sitek2015electron,torres2018conductance,fust2020quantum}.  A nanowire made of a single material, for example InAs, may also become a tubular conductor if the conduction electrons are pushed towards the nanowire walls due to a favorable band bending at the surface \cite{heedt2016adiabatic}.  Assuming a tubular distribution of the electrons in the nanowire, another localization mechanism, that we focus on in is paper, is produced by a magnetic field perpendicular to the nanowire axis, and in that case the electron density within the shell increases in the direction perpendicular
to the field, where the so called snaking states are formed 
\cite{Manolescu13,Rosdahl15,manolescu2016conductance,Chang16}.  

 In a recent paper where two of the present authors where involved, 
it has been predicted theoretically that
a temperature gradient can lead to reversal of thermoelectric current
in tubular nanowires in the presence of transverse magnetic field, at low
temperatures \cite{erlingsson2017}, meaning that the
electrons can either flow from the hot to the cold lead, or vice
versa. This prediction indicates the importance of the magnetic field
effect on the thermoelectric properties of a tubular conductor, but it
still awaits an experimental validation.

 In the present paper we want to address, still theoretically,
the efficiency of a thermoelectric element based on a tubular
conductor in a perpendicular magnetic field.  Efficient
thermoelectric devices are supposed to produce a considerable
electric current, but at the same time to limit the heat flow
\citep{boukai2008silicon,mingo2004thermoelectric}.} These characteristics
are incorporated in the dimensionless figure of merit $ZT$, which is
defined as
\begin{equation}
ZT=\frac{S^2\sigma T}{\kappa} \, ,
\label{zteq}
\end{equation}
where $S$ is the Seebeck coefficient,
$\sigma$ is the electrical conductivity, $\kappa$ the thermal conductivity, and $T$ the temperature. {\blue Hence, there are several parameters that need to be optimized to reach maximum value of $ZT$.  In our physical system we know that the the thermoelectric current is a nontrivial function of the magnetic field and temperature \cite{erlingsson2017}, and the first step of the present paper will be to obtain the Seebeck coefficient. After that,
we will look at the thermal conductivity and finally at $ZT$.}

The thermal conductivity reported for crystalline nanowires
is more than two orders of magnitude lower than the bulk values
\citep{li2003thermal}. Also, the phonon scattering at the nanowire
surface substantially reduces their thermal conductivity and
increases the thermoelectric power factor ($S^2\sigma$)  \citep{wu2013large}.
The diversity of fabrication methods for introducing dopants or
impurities into the lattice is another reason that makes the semiconductor
nanowires important for their thermoelectric characteristics
\citep{pennelli2018thermal,erlingsson2018thermoelectric,dominguez2019nanowires,vuttivorakulchai2018effect,thorgilsson2017thermoelectric}.
The thermal transport in nanoscale systems, whose dimension is much
smaller than the mean free path of electrons, cannot be explained by a
simple law due to the presence of quantum-mechanical features and
strong non-linear behavior \citep{cahill2003nanoscale}. At intermediate
temperatures where ballistic and diffusive phonons coexist, the thermal
conductance decreases non-linearly with the length. And especially at
low frequency, where the acoustic phonons give the major contribution to
the thermal conductance \citep{yadav2006low}. But at low temperatures
charge carriers have an important role in thermal transport 
quantized in multiples of the universal value $\pi^2 k^2_{\rm B}T/3h$,
also electrical conductance is quantized in multiples of universal
value $G_{0}=e^2/h$ which can be understood within Landauer´s formula
\citep{yamamoto2004universal,yamamoto2006nonequilibrium}.

It has been shown that a magnetic field produces large changes in
the thermoelectric properties, including the reduction of thermal
conductivity of charge carriers \citep{wolfe1962effects}. This has
been demonstrated experimentally for Bi$_{88}$Sb$_{12}$ at 78--295\,K
and magnetic fields up to 1.7\,T. The magnetic field has been also
studied for GaAs-Ga$_{1-x}$Al$_{x}$As heterojunction up to 20\,T
\citep{fletcher1986thermoelectric}. The results showed an oscillatory
behavior of thermopower ($S$) with the applied magnetic field. The magnetic
field has been also studied for Bi nanowires array at 50--300\,K
which showed there is an optimum magnetic field for power factor
\citep{hasegawa2005electronic}. 


{\blue The material of the paper is structured in these steps: In Section
2 we present the model and the energy spectra of our system, the tubular
conductor in perpendicular magnetic field.  Then, in Section 3 we discuss and show
the results for the Seebeck coefficient, for the thermal conductivity, and
for the figure of merit.  Finally, the conclusions are collected in Section 4.}

\section{Model and methods}

In this paper, we consider electronic transport in a tubular, cylindrical
nanowire, in the presence of a longitudinal temperature difference and a uniform
magnetic field transverse to the axis of the cylinder. The conduction
takes place only in a narrow shell at the surface and not through the
bulk \citep{heedt2016adiabatic}.

The Hamiltonian of the system can be written as
\begin{equation}
    H=\frac{(-i \hbar \nabla+e{\bm A})^2}{2m_{\rm eff}}-g_{\rm eff} \mu_{B} s B
    \label{hamilt}
\end{equation}
where $B$ is the magnetic field {\blue in the $x$ direction, i.e. perpendicular 
to the nanowire length which is oriented along the $z$ axis, and
${\bm A = (0,0,By)}$ is the} corresponding vector potential. Also, $e$ is the electron charge,
$m_{\rm eff}=0.023m_0$ and $g_{\rm eff}=-14.9$ are the effective electron
mass and bulk g-factor of InAs, $\mu_{B}$ is the Bohr's magneton {\blue and
 $s = \pm 1$ is the spin label}. {\blue We chose the effective mass and g-factor as for InAs because this is a relatively common material used for core-shell nanowires.} We assume that electrons propagate along the nanowire without interacting with other electrons.

System parameters are $R=30$\,nm, $B=0-3$\,T and also
by considering material parameter for InAs, we can calculate the heat
current and electrical current driven by the temperature bias and chemical
potential difference between the two ends of the nanowire, where we assume
external leads are contacted.  We calculate the charge current $I_c$
and heat current $I_Q$ through the nanowire using the Landauer approach:
\begin{equation}
    I_{c}=\frac{e}{h} \int {\cal T}(E)[f_{R}(E)-f_{L}(E)]dE,
    \label{eq2}
\end{equation}
\begin{equation}
    I_{Q}=\frac{1}{h} \int {\cal T}(E)[E-\mu][f_{R}(E)-f_{L}(E)]dE,
    \label{eq3}
\end{equation}
where ${\cal T}$ is the transmission function, and
\begin{equation*}
    f_{L,R}(E)=\frac{1}{1-e^{(E-\mu_{L,R})/kT_{L,R}}}
\end{equation*}
is the Fermi function for the left (L) or right (R) reservoir with
chemical potential $\mu_{L,R}$ and temperature $T_{L,R}$. We consider a
temperature bias  $\Delta T=T_{R}-T_{L}> 0$, with $T_{L}$ always
fixed at $0.5$\,K, and a chemical potential bias $\Delta \mu = \mu_{R}-\mu_{L}$,
with $\mu_{L}=\mu -\Delta \mu /2$ and $\mu_{R}=\mu +\Delta \mu /2$, where $\mu$
is fixed at $4.2$\,meV and $\Delta \mu$ is varied between $0-0.4$\,meV.

{\blue Ballistic transport of electrons in nanowires leads to a transmission ${\cal T}$, as a function of energy, with a step behavior. Nanowires showing such step-like behaviour have been measured, and in the presence of a low, but achievable impurity density the steps are still visible \citep{kammhuber2016conductance}. 
Based on this experimental fact we assume ballistic transport in our system.
} The transmission function in the presence
of impurities can be obtained with the recursive Green's function method
\citep{ferry1999transport,erlingsson2017}. For nanowires with inhomogeneities, such as impurities, a nonuniform diameter, surface changes, or stacking faults, the conductance becomes a series of transmission resonances due to quantum dot like states \citep{wu2013large}. In that case transport calculations based on elastic scattering have been performed up to
24\,K, so we can neglect inelastic collisions in our temperature range
\citep{wu2013large,erlingsson2017,erlingsson2018thermoelectric}.

\begin{figure}[h]
    \centering
    \includegraphics[width=1.06\linewidth]{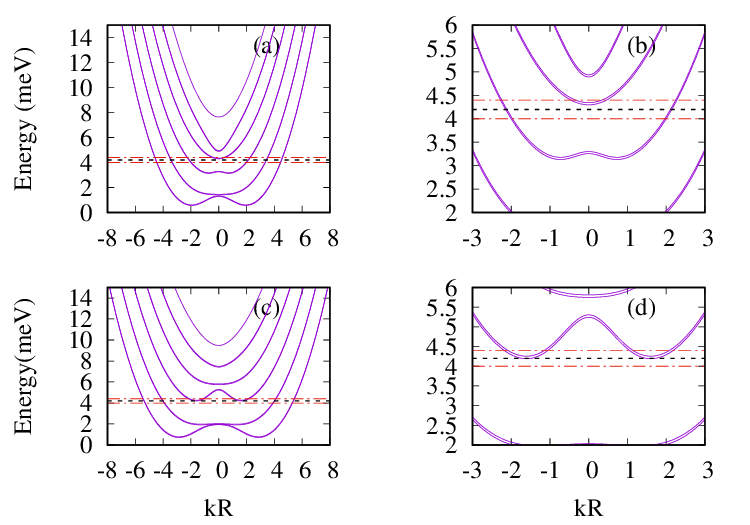}
    \caption{Energy spectra for a cylinder of infinite length and radius $R=30$\,nm in presence of transverse magnetic field $B=1.8$\,T (a) and (b), and $B=2.5$\,T (c) and (d). The black horizontal dotted lines indicate the chemical potential $\mu=4.2 $\,meV and red horizontal dotted lines indicate the chemical potential $\mu=4.0 $\,meV $\&$ $\mu=4.4 $\,meV, respectively. (b) and (d) are the magnified image of (a) and (c) around the $\mu$.}
    \label{band}
\end{figure}

{\blue The eigenstates of the Hamiltonian (\ref{hamilt}) are calculated
by numerical diagonalization in a basis set corresponding to plane waves
$\exp(ikz)$, with $k$ the wave vector in the longitudinal direction, 
and angular momentum eigenfunction
$\exp(im\varphi)$, with $m=0,\pm 1,\pm 2,...$, in the transversal plane $(x,y)$
where the electrons are confined on a circle of radius $R$
\cite{Manolescu13}. The resulting energy spectra for magnetic fields $B=1.8$\,T
and 2.5\,T are shown in Fig.\,\ref{band}.  These spectra have an
interesting feature: they may not always monotonic functions of $k$ when
it has a fixed sign.  Meaning that the transport channels, i.e. the number of
states with a fixed energy, which in general increases with increasing the
energy, in this case may also decrease.  Consequently, the thermoelectric
current may change sign \cite{erlingsson2017}}.

%

In the transport calculations we will consider that only electrons
that are close to $\mu$ in energy contribute to the heat transport. The
chemical potentials are chosen such that $\mu$ is close to a subband
bottom.  In Fig.\,\ref{band} (b) and (d) we show this energy interval
for two values of the magnetic field.

\vspace{2cm}

\section{Results and discussion}

\subsection{Seebeck coefficient} 

The Seebeck coefficient or the thermopower, $S$, is
defined by the ratio $-\Delta V/ \Delta T$, where the voltage difference
$\Delta V$ is produced in presence of a small temperature difference
$\Delta T$ between two points of a conductor, under an open circuit
condition. Usually the thermopower consists of two additive contributions:
diffusion $S^d$, and phonon drag $S^g$. The first one originates from
the diffusion of carriers (electrons or holes) and second one comes
from the momentum that is transferred to carriers via their coupling to
non-equilibrium acoustic phonons in the presence of a temperature gradient
\citep{tsaousidou2010thermopower,fletcher1997thermoelectric,fletcher1994thermoelectric}.
For the total thermopower $S=S^d+S^g$ there is a very good agreement
between theory and experiments at temperature below $21$\,K for example 
in bulk silicon \citep{behnen1990quantitative}.  However,
at this low temperatures, where normally $S^d$ dominates, we can neglect
the phonon drag contribution to the non-equilibrium electron distribution
function, and consider elastic scattering as the main mechanism that
limits the electrons' momentum relaxation time. 

The Seebeck coefficient is important for two 
reasons. First, this coefficient provides fundamental information about
the electronic energy structure and the electron scattering mechanism in
a system. Second, there is some evidence suggesting that thermoelectric
energy conversion can be more efficient in low-dimensional systems
\citep{tsaousidou2010thermopower2}.
For example, for a semiconductor, a large magnitude of Seebeck
coefficient requires only one type of carrier, because mixed n-type and
p-type conduction will send both carriers through contacts, leading to a
reduced Seebeck voltage.  The relation between the carrier concentration
and Seebeck coefficient, for bulk states, is given by:

\begin{equation}
    S=\frac{8\pi^2k_{B}^2}{3eh^2} m^* T  \left( \frac{\pi}{3n}\right)^{2/3} \ ,
    \label{equation6}
\end{equation}
where $n$ is the carrier concentration and $m^*$ is the effective mass of carriers. Although a low carrier concentration of insulators and semiconductors result in large Seebeck coefficient, it also leads to a low electrical conductivity, $1/\rho=\sigma=ne\mu_c$, where the electrical conductivity and electrical resistivity are related to $n$ through the carrier mobility $\mu_c$ \citep{snyder2011,hicks1993thermoelectric,rowe2018crc}. There is also another conflict with the effective mass of the charge carriers, in a manner that large effective mass provide high thermopower but low electrical conductivity.

\begin{figure}[h]
    \centering
    \includegraphics[width=1.0\linewidth]{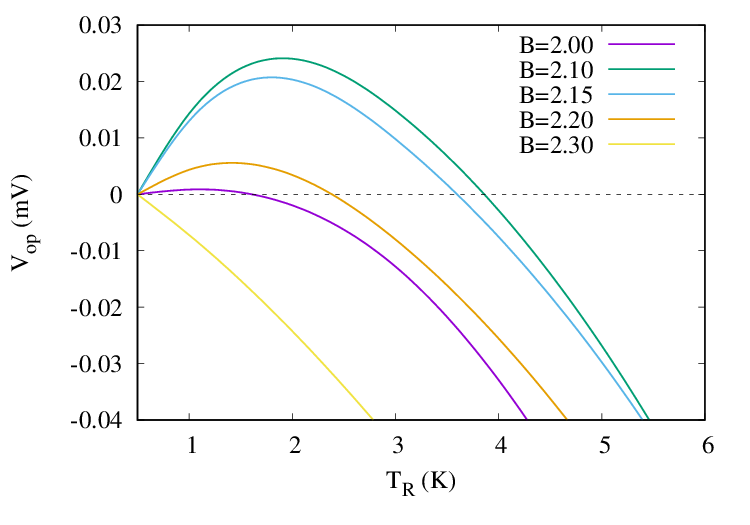}
    \caption{ Open circuit voltage as function of right lead temperature in indicated values of magnetic field}
    \label{open circuit voltage}
\end{figure}

{\blue In the present work we consider ballistic transport
such that the electronic energy spectra have the main role
in the behavior of the thermopower.  We assume that the
scattering due to impurities have negligible effects, and that
is a reasonable approximation in a sufficiently clean system
\cite{thorgilsson2017thermoelectric,erlingsson2018thermoelectric}}.
Using the transmission functions for the calculated energy spectra
we determine the voltage in an open circuit condition, $V_{\rm op}$,
from the chemical potential bias necessary to bring to zero the electric
current in the system [Eq.\,(\ref{eq2})]. That means we evaluate $\Delta
\mu=\mu_{R}-\mu_{L}=eV_{\rm op}$ as a function of the temperature of the
right lead, for different values of magnetic field.  {\blue As one can
see in Fig.\,\ref{open circuit voltage} the open circuit voltage has a nonlinear 
dependence on the temperature bias, for magnetic fields between $2.0-2.3$\,T. More
remarkably is though the change of sign as a function of the temperature, which 
occurs because of the nonmonotonic variation of the transmission function with respect 
to the energy \cite{erlingsson2017}.}

We obtain numerically the Seebeck coefficient, $S=V_{\rm op}/\Delta
T$, {\blue as the linear coefficient of $V_{\rm op}$ as function of the
temperature gradient, $S=V_{\rm op}/\Delta T$, by performing the procedure
described above with a small temperature bias, $\Delta T=T_{R}-T_{L} =
0.1$\,K. This time both $T_R$ and $T_L$ are varied.}
We calculate the Seebeck coefficient at the specific $\mu$ situated
close to a subband minimum, where {\blue the insensitivity of $S^g$ to 
the energy dependence of electron relaxation time has a direct impact on 
the phonon-drag contribution to the magnetothermopower tensor, that results in } 
$S^d$ becomes dominating over $S^g$.  
\begin{figure}[h]
    \centering
    \includegraphics[width=1.0\linewidth]{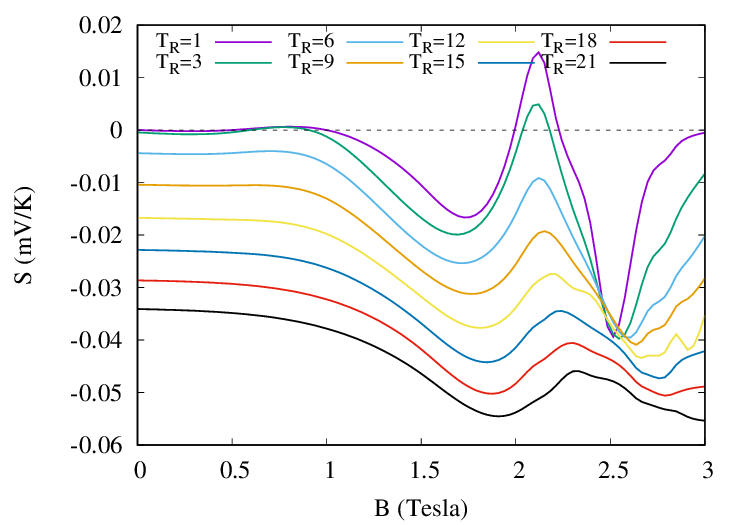}
    \caption{Seebeck coefficient with magnetic field for different values of $T_{R}$ where $\mu=4.2$\,meV and $T_L=T_R-0.1$\,K (i.e. $T_L \approx T_R$). }
    \label{Seebeck coefficient}
\end{figure}

Fig.\,\ref{Seebeck coefficient} shows the variation of $S$ with the magnetic field at
different temperatures $T_R \approx T_L$. It can be clearly seen that $S$ starts with
nearly a constant value, and then continues with an oscillatory behavior,
with increasing magnetic field, for all $T_R$ studied here. The amplitude
of oscillation becomes smaller at higher $T_R$. An oscillatory behavior
with respect to magnetic field has been obtained earlier in the 2D
electron gas, in the fractional quantum Hall regime, but without a sign change 
\citep{fletcher1986thermoelectric}. {\blue Oscillations of the thermopower vs. 
the chemical potential with sign changes were predicted a long time ago for 
quantum dots \cite{Beenakker92}, and confirmed experimentally
\cite{Svensson12}, but as a consequence of isolated resonances.}
Besides, at low fields, i.e. in the constant regime of $S$, it has
been shown that the increase of the temperature gradient increases $S$,
and that is in agreement with our results.  Indeed, at low temperatures
(1--6\,K) and magnetic field above 2\,T our oscillations and the sign change 
of the thermopower are expected from the similar behavior of the
thermoelectric current and open circuit voltage discussed above. 
And at higher temperatures the oscillations become smoother and without a sign change.


\subsection{Thermal conductivity} 

\begin{figure}[h]
    \centering
    \includegraphics[width=1\linewidth]{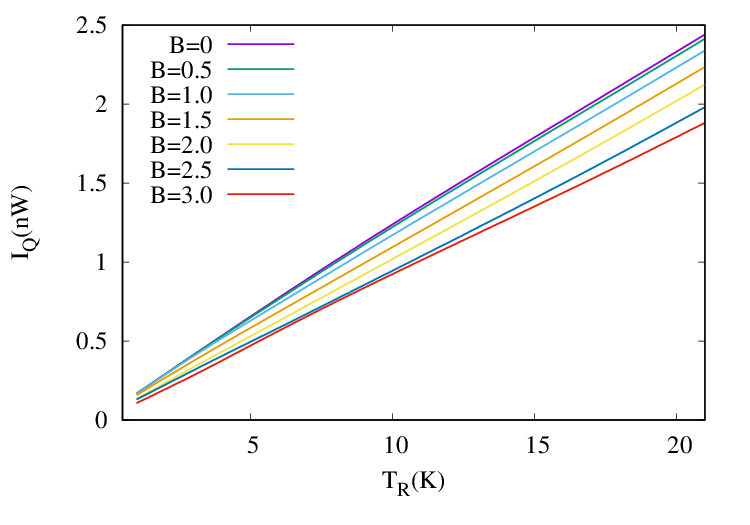}
    \caption{Heat current as function of the temperature of the right contact $T_{R}$ for the indicated magnetic field values and the chemical potential $\mu=4.2$\,meV}
    \label{heat current}
\end{figure}

One of the fundamental factors for high efficient thermoelectric
conversion is the thermal conductivity, which needs to be minimized. 
There are many possibilities to
reduce the thermal conductance of a nanosystem \citep{li2003thermal}.
{\blue Our next step is to evaluate the heat transported by the electrons of
our system, which accompany the transport of electric charge.} 
We calculate the heat current as function of the temperature of the right
lead, for different values of the magnetic field, using Eq.\,(\ref{eq3}), and $\Delta \mu=0$. 
We can see in Fig.\,\ref{heat current} that with increasing the magnetic
field strength, the heat current decreases, but not dramatically. Also,
the heat current is more influenced by the magnetic field at high temperatures,
for example at $20$\,K compared to $0.5$\,K. The reason of this behavior
is the distribution of carriers over the energy states 
such that more electrons are localized due to the closed cyclotron motion imposed
by the field, and fewer of them are available for transport.
{\blue Note also that the heat current does not change sign, as the charge current does, that is in agreement with the second law of thermodynamics.}

\begin{figure}[h]
    \centering
    \includegraphics[width=1.0\linewidth]{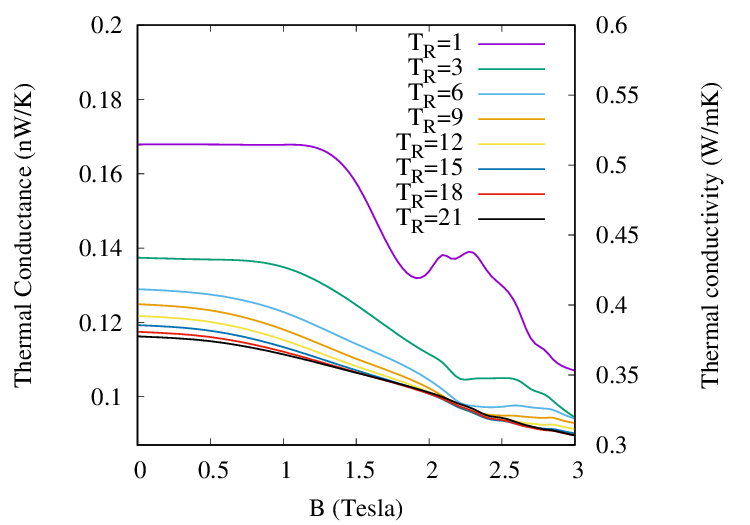}
    \caption{Thermal conductivity and thermal conductance of the tubular nanowire vs. magnetic field for several indicated temperatures of the right lead.  The nanowire length is $L=100$\,nm and radius $R=30$\,nm.}
    \label{thermal conductance}
\end{figure}

The thermal conductance, $\bar{\kappa}=dI_Q/dT_R$, and thermal conductivity, $\kappa=L/(\pi R^2)\bar{\kappa}$, as functions of the magnetic field, for different temperatures of the right lead, are shown in  Fig.\,\ref{thermal conductance}. 
{\blue Here we use the whole cross sectional area of the cylinder, although the 
transport of both heat and charge occurs through the shell defined by the outer 
surface.  The full cross sectional area is however relevant if the core is 
also included in the heat transport with phonons, which are neglected at our 
low temperatures.}  

The figure clearly indicates two regions with (i) constant $\kappa$
and $\bar{\kappa}$ at low fields and (ii) a non-linear reduction at
higher applied magnetic field.  We can see almost the same trend for
$\kappa$ (or $\bar{\kappa}$) for different temperatures, but that is
more evident at lower $T_R$.  {\blue It is obvious that the increase of
the magnetic field leads to a reduction of the thermal conductance and
thermal conductivity, but the amount of these changes are different
for each temperature.} The reduction of the contributions of charge
carriers (electrons or holes) to the thermal conductivity has been
observed in experimental studies for both bulk and nanowire arrays
\citep{wolfe1962effects,hasegawa2005electronic}.

\subsection{Figure of merit} 

\begin{figure}[h]
    \centering
    \includegraphics[width=1.0\linewidth]{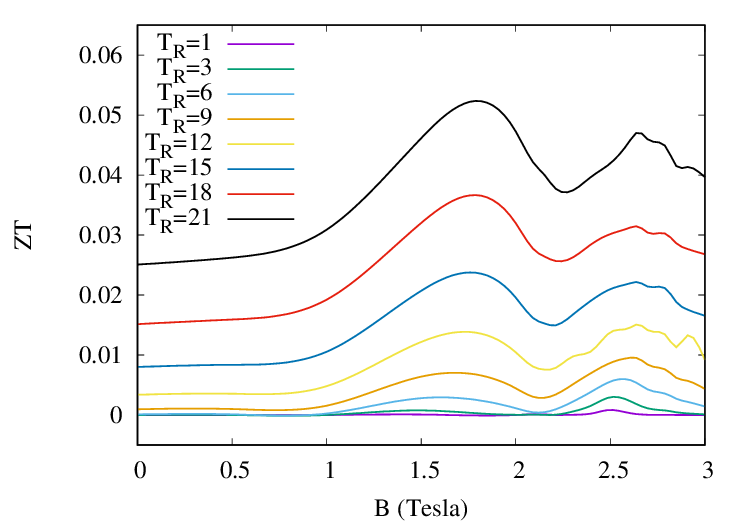}
    \caption{$ZT$ as function of magnetic field where $\mu=4.2$\,meV.}
    \label{zt}
\end{figure}

Next, in order to find the optimum conditions for thermoelectric
conversion, we need to calculate $ZT$ using Eq.\ (\ref{zteq}). 
High electrical conductivity and low thermal conductivity is required {\blue to maximize or optimize} $ZT$, that is achievable by considering lattice thermal conductance and materials characteristics \citep{dresselhaus2007new,chen2003recent,rosi1968thermoelectricity}. {\blue In $ZT$ formula the cross sectional area of the full cylinder
($\pi R^2$) used in the thermal conductivity is going to be compensate by
the area used in electrical conductivity. Electrical conductivity
 calculated from $\sigma= dI_{c}/dV$, where for each specific temperature and magnetic field several values of $I_{c}$ and $V=(\mu_R-\mu_L)/e$ were calculated separately and got differentiated.}
Fig.\,\ref{zt} represents $ZT$ as function of magnetic field for
different temperatures of the right contact. 

Again, the figure shows two
regions, of constant and oscillating $ZT$, respectively. It is also
clear that increasing the temperature shifts $ZT$ to higher values,
in both regions, non-linearly. Also, at low temperatures, the figure
of merit shows limited changes with increasing magnetic field, but
for temperatures more than 9\,K we have some obvious variation of $ZT$
values. There are two peaks for $ZT$ in the figure, which is a specific
result of magnetic field presence. At all temperatures studied here,
the peaks are located at about $B=1.8$ T and 2.5 T with a slight shift to
higher field at higher temperature. Thus, regardless of the temperature
difference, there is an optimum magnetic field that leads to maximum $ZT$. 
{\blue In addition, although we are referring to low temperatures,
doubling the value of $ZT$ just by applying to the system an external 
magnetic field is interesting tuning possibility.}

{\blue With increasing the temperature above 20\,K, one would expect an increase 
of $ZT$ simply because of the temperature factor in the definition,
Eq. (\ref{zteq}).  But, of course, the phononic contribution to the heat transport
increases with temperature, and the phonon drag and lattice vibrations 
will end up by dominating over
the diffusive heat transport due to electrons.
However, experimental values for the thermal conductivity  $\kappa$
in nanowires with diameters between 20-100\,nm shows a saturation behavior 
for temperatures above 100\,K to values like 10-40 W/mK \citep{li2003thermal}. 
Therefore, for such a temperature we can expect the figure of merit of 
our system to become roughly ten times bigger than the values shown in
Fig.\ \ref{zt}.}

\section{Conclusions}

In this paper we have calculated the most important thermoelectric
parameters, such as heat and electric current, the open circuit voltage
$V_{\rm oc}$, Seebeck coefficient $S$, thermal conductivity $\kappa$,
and figure of merit $ZT$, produced by electrons confined within a tubular
nanowire due to a temperature bias, in presence of transverse magnetic
field. To this end, heat current and electrical current variations are
obtained in the temperature range between ($0-20$\,K).  Increasing the
magnetic field leads to reduction  in thermal conductivity, which is
more pronounced at lower temperatures.
{\blue The energy spectrum of electrons is a nonmonotonic function of 
the wave vector along the nanowire, and so is the transmission function with respect to 
the energy.  Consequently $V_{\rm oc}$ can change sign when the temperature gradient 
or the magnetic field increase.}  Both $S$ and $ZT$  have a constant to oscillatory
transition with increasing the magnetic field. For example for a
cylinder radius of 30\,nm, $ZT$ presents two peaks at about 1.8 and 2.5\,T
which are independent of the temperature.   

{\blue These feature allow a substantial tuning of the thermoelectric
response of the nanowire with changing the temperature or with applying
an external magnetic field. To the best of our knowledge, although
several groups have performed thermoelectric measurements of nanowires
at low temperatures 
\cite{prete2019thermoelectric,diez2020enhanced,fust2020quantum,li2003thermal,wu2013large,pennelli2018thermal}, 
experimental investigations of tubular conductors
based on core-shell geometry in a transverse magnetic field have not
been reported yet. Therefore it is our hope that our theoretical results
will stimulate such an experimental work.}

\begin{acknowledgments}
This work was supported by the Icelandic Research Fund, Grant 195943-051.
\end{acknowledgments}

\bibliography{Bibliography2}

\end{document}